\documentstyle[12pt]{article}
\begin{document}

\baselineskip=24pt \centerline {\large {\bf On the Particles Transport}} 
\centerline {\large {\bf Between Embedded Clusters}} 
\centerline{\bf Florin Despa } \baselineskip=18pt 
\centerline{\sl Department of Theoretical Physics,} 
\centerline{\sl Institute of Atomic Physics,} 
\centerline{\sl POBox  MG - 6, M\u{a}gurele, Bucharest, 
R - 76900, Rom\^{a}nia}
\centerline{\sl e-mail: despa @ roifa. if. ro}
\centerline{\sl Fax: 40. 1. 312. 2247}
\baselineskip=24pt \centerline{\bf Vladimir Topa } \baselineskip=18pt 
\centerline{\sl Institute of Physics and Technology of Materials,} 
\centerline{\sl POBox  MG - 7, M\u{a}gurele, 
Bucharest, R - 76900, Rom\^{a}nia}

\vspace{ 24pt} \baselineskip=24pt

\hspace{0.5cm}{\bf Abstract:} Diffusional interactions within the Ostwald
Ripening process are analyzed in the present paper. An {\it off-centre
diffusion approach \/} is performed to point out the direct correlation
between the size of clusters. Herein the diffusion solution is derived as a
function of both the growing and shrinking cluster sizes. Also, it is shown
that the frequency transfer of particles between the shrinking cluster and
the growing one may acquire high values due to the medium polarization. As a
result, the temporal power law of the cluster growth derived in this
theoretical model differs from that predicted by the {\it LSW \/} theory.
Experimental data for $Ag$ clusters embedded in a $KCl$ matrix are analyzed
both by the present theory and by the $LSW$ theory.\\ \vspace{ 36pt} 
\baselineskip=24pt

\newpage

\topmargin=-0.5in \baselineskip=24pt

\section{Introduction.}

\indent As it is well known, one can distinguish two regimes in the phase
separation process following the formation of nuclei. The initial stage is
the growth of these nuclei by the condensation of solute on their surface.
The second stage is known as the {\it Ostwald Ripening (OR) \/} process
where the particles flow from shrinking clusters to growing ones.

The kinetics of domain growth in the late stages of diffusion-limited
spinodal decomposition {\it Ostwald Ripening (OR) \/} have been studied by a
variety of methods \cite{1,2,3,4,5,6,7,8}. 
The {\it Lifshitz, Slyozov and Wagner (LSW) \/} theory predicts that the
average droplet radius $R$ grows with time $t$ as $R(t)=\Gamma t^\alpha $,
where $\Gamma =const.$ and $\alpha =1/3$ and that the distribution of
droplet sizes reaches a material-independent universal form when properly
scaled. Most simulations and experiments measure $\alpha $ exponents in the
range $0.15$ to $0.25$ \cite{9,10,11,12,13,14,15,16,17,18,19,20,21,22}.
below the theoretical value $\alpha =1/3$. Moreover, the measured size
distributions are typically broader than the $LSW$ prediction. This
discrepancy has been attributed either to diffusion effects at an interface
or the inadequacy of mean field description for the systems or to
insufficiently long simulation times. Several authors have developed
improved theoretical models that take into account interaction 
effects \cite{3}.
These models involve expansions in power of the parameter $\sqrt{\phi }$
(where $\phi $ is the volume fraction of the minority phase), whose
importance was first recognized by {\it Tokuyama \/} and {\it Kawasaki \/} 
\cite{23}. To first order in $\sqrt{\phi }$, interactions give rise to two
types of corrections: a direct correlations between droplet pairs where
small droplets are likely to be surrounded by large ones as well as a
''medium polarization'' in which the rate of evolution of a droplet is not
only a function of its radius $R$, but also of the droplets within a
neighborhood of size $\xi $. The models reproduce the broadening of
experimental distributions while predicting that correlations do not alter
the value of the $LSW$ exponent $\alpha $ just as observed in experiments.
Recent experimental results \cite{24} have revealed the above types of
correlation effects in the two-dimensional coarsening process. Thus,
according to the direct correlation effect small (large) cluster are more
likely to be found near large (small) ones . The other correlation effect,
experimentally observed, is a medium polarization according to which the
rate of change of the size cluster is determined not only by its size but
also by the influence of others in its surroundings. Thus, the medium
polarization around two nearby clusters promotes the accelerated shrinkage
of one and growth of the other, the rates of shrinkage and growth being
larger than if both clusters were isolated. Taking into account these
correlation effects it has been proven \cite{25} that within an {\it %
off-centre diffusion \/} approach \cite{26,27} the temporal law of the $OR$
process in two-dimensions can be, under some circumstances, different from
that predicted by the $LSW$ theory.

The purpose of this paper is an examination of the cluster growth where the
dynamic is dominated by the coarsening process. As in the previous case of
the $OR$ in two-dimensions \cite{25}, we will suppose that the particles
transport from shrinking cluster to growing one occurs by an {\it off-centre
diffusion \/} mechanism \cite{26,27}. The identifying of the {\it Markowian
chains \/} (as within the {\it Flux over Population Method \/} \cite{28}) is
based on a local feature of the medium, according with, the small cluster
(which disappears during the $OR$ process) is likely to be surrounded by the
greater ones (which increase by the incorporation of mass into them). In
this way, the diffusion solution will be determined as a function of both
the growing and shrinking cluster sizes. Also, it is shown that the
frequency transfer of particles between the shrinking cluster and the
growing one may acquire high values due to the medium polarization.
Particular properties of the clusters are included in the model. As a
result, the temporal power law of the cluster growth derived in this
theoretical model differs again, as in the two-dimensional case, from that
predicted by the {\it LSW \/} theory. Some experimental results on the
growth of the $Ag$ clusters embedded in a $KCl$ matrix will be analyzed by
the present theory.

It must be pointed out that the present approach of the three-dimensional 
$OR
$ process can work only under assumption that the correlation effects occur
(especially, the direct correlation between the cluster sizes). This fact
can exist if the clusters nucleate and grow at dislocation lines and/or at
grain boundaries (situation which is frequently supposed for clusters
embedded in solid matrix \cite{29}).

Far to elucidate the controversy regarding the general theory of the $OR$
process (especially related to the famous $\frac 13$), the {\it off-centre
diffusion \/} approach of $OR$ gives, at least, a real way to account for
the correlation effects. Moreover, this distinguishes from the others in
this branch by the fact that the clusters act as entities in theirself and,
consequently, the temporal power law of $OR$ is derived in connection with
their particular properties \cite{25}.

\topmargin=-0.5in \baselineskip=24pt

\section{Off-centre Diffusion Approach of the $OR$ Process.}

\indent 
As we have said in the previous section, according to the direct correlation
effect within the two -dimensional $OR$ process experimentally observed by 
{\it Krichevsky\/} and {\it Stavans \/} \cite{24}, small (large) clusters
are more likely to be found near large (small) ones . Supposing that this
fact is appropriate also in three-dimensions, let us examine the effect of
such a correlation on the diffusion process in the $OR$ phenomenon.
Consequently, we will assume the situation which is shown in Fig. 1, where
there exist a cluster of $N$ sites slightly displaced around the position of
the large cluster and, only $N_o$ sites ($N_o<N$) around the position of the
small cluster. These off-centre sites (the ''kinks'' of the cluster surface)
serve both to the particles motion on the cluster surface \cite{30} and as
available sites for the particles transfer from the shrinking (small)
cluster to growing (large) cluster \cite{25,26,27}. In the following, we
assume that the motion among the available sites of the same cluster is
described by the frequency $p_o$ while the transfer frequency from shrinking
(small) cluster to growing (large) one is $p$, $p_o\gg p$. In this way, the
particles leave the ''kink'' sites of the surface of the shrinking cluster
and ''condense'' in the ''kink'' sites of the nearest-neighbour growing
cluster. The equations for the $N_o$ concentrations $n_i(x,t)$, 
$i=\overline{
1,N_o}$, can readily be written as: 
\begin{eqnarray}
& &\frac{\partial n_1}{\partial t} = p_o \sum_{i=1}^{N_o} (n_i - n_1) +
p \sum_{j=1}^{N} \left( n_j (x + \xi) - n_1 \right) \\ \nonumber
& &\vdots \\ \nonumber
& &\frac{\partial n_{N_o}}{\partial t} = p_o \sum_{i=1}^{N_o} (n_i - 
n_{N_o}) +
p \sum_{j=1}^{N} \left( n_j (x + \xi) - n_{N_o} \right) . \nonumber
\end{eqnarray}
We perform the power-series expansion of the concentration function in (1).
Since, as usual in diffusion processes, we are interested only in slowly
varying functions in time and space \cite{26,27}, we may neglect the terms
containing the {\it first-order \/} derivatives in the series expansion of
(1). Equations (1) can, therefore, be approximated by 
\begin{eqnarray}
& &\frac{\partial n_1}{\partial t} = p_o \sum_{i=1}^{N_o} (n_i - n_1) +
N \frac{1}{2} p \xi^2 \frac{\partial^2 n_1}{\partial x^2} \\ \nonumber
& &\vdots \\ \nonumber
& &\frac{\partial n_{N_o}}{\partial t} = p_o \sum_{i=1}^{N_o} (n_i - n_{N_o}) 
+
N \frac{1}{2} p \xi^2 \frac{\partial^2 n_{N_o}}{\partial x^2} , \nonumber
\end{eqnarray}
whose Fourier transforms read 
\begin{eqnarray}
& &n_{1q} \left(\omega - N\frac{1}{2}p\xi^2q^2 \right) + n_{2q}p_o + \ldots 
+
n_{N_oq}p_o = 0 \\ \nonumber
& &\vdots \\ \nonumber
& &n_{1q}p_o + \ldots + n_{(N_o - 1)q}p_o + n_{N_oq} \left(\omega - 
N\frac{1}{2}p\xi^2q^2 \right) = 0 . \nonumber
\end{eqnarray}
Looking at the system of equations (2) and at their {\it Fourier transforms
\/} we can see that the problem amounts to finding the lowest eigenvalue of
a system of equations which has the general matrix form \cite{27}, 
\begin{equation}
A=p_oA_o+pA_1+A_2.
\end{equation}
Here, $A_o$ describes the diffusion among the off-centre sites belonging to
the same cluster; $A_1$ corresponds to the {\it second-order \/} expansion
of the concentration functions; and $A_2$ includes the {\it higher-order \/}
contributions of the derivatives. We note that in the long-wavelength limit 
$A_2$ vanishes. The lowest eigenvalue can be obtained by a 
perturbation \cite{27} 
given by 
\begin{equation}
\omega =\overline{n}A_1n=\frac 12Np\xi ^2q^2,
\end{equation}
where $n$ is the (column) vector adjoint to eigenvector 
\begin{equation}
\overline{n}=N^{\frac{-1}2}\left( 1,1,\cdots ,1\right) .
\end{equation}
The diffusion solution is given by 
\begin{equation}
n(x,t)=\frac{n_0}{\sqrt{2\pi Np\xi ^2t}}\cdot \exp {(-\frac{x^2}{2\cdot
Np\xi ^2t})}.
\end{equation}

This equation gives the particles number per unit length at the time $t$ and
at the distance $x$ due to the diffusion of an initial $\delta $ - form
concentration of particles. $N$ is in direct proportion with the cluster
surface and, in a crude approximation, can be expressed by 
\begin{equation}
N\approx \frac 12\cdot \frac{R^2}{a_o^2},
\end{equation}
where $a_o$ is the atom radius and $R$ the radius of the growing cluster.
The above equation establishes that, due to the geometrical obstructions,
see Fig. 1, only half of the peripheral off-centre sites (the ''kinks'') are
available to receive diffusing particles.

The diffusing particles come, in the $OR$ process, from shrinking cluster
and, indeed, we must take into account its dissociation rate. As it is well
known, the cohesive energy per atom decreases with decreasing cluster size 
\cite{31} and, therefore, the dissociation rate for shrinking clusters
become considerable greater in comparison with growing ones. This is
important to gain physical insight in the $OR$ process. The shrinking or
growing of an cluster begins from a critical radius that depends on its
size. A common definition of the critical radius states that it is the
radius of a droplet which is instantaneously neither growing nor shrinking.
The dissociation rate is related, in the $RRK$ theory {\it (Rice,
Ramsperger, Kassel) \/} \cite{32}, to both the thermal energy $E_o=E_o(T)$ ($
T$ stands for the temperature) and the dissociation energy of the particle $
E_D$, 
\begin{equation}
K(T)=\nu \left[ \frac{E_o-E_D}{E_o}\right] ^{s-1}\,.
\end{equation}
Here, $\nu $ is the vibrational frequency, $s$ is the number of vibrational
degrees of freedom of the cluster and $T$ is the temperature. One seems that
the excitation of the cluster, ultimately, causes heating and dissociation
and that to a large extent the excitation mechanism is decoupled from the
dissociation. Thus, with such a simplification, the dissociation rate can be
calculated by (9). In this way, we may find the total amount of dissociated
particles from the shrinking (small) cluster during the thermal annealing as 
\begin{equation}
n_o=W\cdot K(T)\cdot t,
\end{equation}
where $W$ accounts for the surface atoms of the cluster and $t$ is the time
of the thermal annealing. Further, the $n_o$ entering in equation (7) is
replaced by the above amount.

The other correlation effect, theoretically assumed in \cite{23} and
experimentally observed in two-dimensional $OR$ \cite{24}, is a medium
polarization according to which the rate of change of the size cluster is
determined not only by its size but also by the influence of others in its
surroundings. Thus, the medium polarization around two nearby clusters
promotes the accelerated shrinkage of one and growth of the other, the rates
of shrinkage and growth being larger than if both clusters were isolated.
The medium polarization is due to the electrostatic interaction between the
charges associated with each shrinking (negative charge) or growing
(positive charge) cluster. This charge is proportionally to the rate of
change of the cluster area. The medium polarization consists in the
appearance of an electrostatic potential 
\begin{equation}
\Phi (r) = C \frac{e^{-\frac{r}{D}}}{r}, 
\end{equation}
as solution of the {\it Poisson - Boltzmann equation \/} \cite{33}. $C$ is a
constant depending on the cluster size and $D$ is the {\it Debye length \/}.
The {\it Debye length \/} is in inverse proportion with $\sqrt {M}$ where $M$
is the number of clusters within the neighborhood of the reference cluster
(the shrinking or growing cluster). Indeed, the activation energy for
particles transfer from shrinking cluster to growing one is considerably
lowered due to this electrostatic potential ($\Phi $), thereby enhancing the
transfer frequency (see eq. 7) 
\begin{equation}
p = \nu \exp{\left(-\beta \left( E_b - e \Phi \right) \right)}, 
\end{equation}
where $E_b$ denotes the threshold energy for activation, $\nu$ is a
prefactor, and $\beta = \left(K_BT \right)^{-1}$. In this way the medium
polarization accelerates both the shrinkage of a small cluster and the
growth of a large one. As we have said in the introduction part, this
correlation effect (as well as the former) can be properly understood for
embedded clusters in solid matrix only if the nucleation sites, which really
promote the growth of clusters, occur at dislocation lines or/and at grain
boundaries \cite{29}.

Diffusing particles added to a growing cluster having an initial critical
radius $R_o $ leads to an increase in its radius to $R$; 
\begin{equation}
n = \rho \frac{4 \pi}{3} \left( R^3 - R_{o}^{3} \right) , 
\end{equation}
where $\rho$ is the particles concentration in the cluster. Also, taking
into account (7) we can express $n$ by 
\begin{equation}
\int^{R}_{R_o} n\left( \xi - R, t\right) dR  = \frac{WK(T)}{\sqrt{\frac{\pi 
}{a_{o}^{2}} p\xi^2 }} \sqrt{t} \cdot \int^{R}_{R_o} \frac {1}{R} \exp{%
\left(-\frac{(\xi - R) ^2}{\frac{1}{a_{o}^{2}}p\xi^2t} \right)} dR , 
\end{equation}
where $\xi$ stands for the separatrix between the shrinking cluster and the
growing one. For large $t$ and as $\xi \approx R $ \cite{24} the exponential
function vanishes and the above equation becomes 
\begin{equation}
\frac{ R^3 - R_{o}^{3}}{\ln {\frac{R}{R_o}}} = \frac{WK(T)a_o}{\rho \sqrt{
\pi^3 p\xi^2 }} \sqrt{t}. 
\end{equation}
The last equation gives the time $t$ for an increase of the cluster radius
from a radius $R_o$ to $R$ by an {\it off-centre diffusion mechanism \/}.
Indeed, the diffusion process is related to the frequency transfer $p$ and,
as we have said, within the particular dependencies of $p$ we must take into
account (11).

\topmargin=-0.5in \baselineskip=24pt

\section{Experimental.}

\indent Metal clusters can be produced with ease in solid matrix \cite{29}.
For example, electrolytical or additive colouring of the alkali halide
crystals containing relatively high impurity concentration 
($ \approx 10^{18}$
impurities per $cm^3$) lead, directly,
to cluster formation \cite{29,35,36}. Another, more adequate, method in 
order
to study the kinetic aspects of the embedded clusters is the
 thermal annealing of alkali halide crystals containing negative
metallic ions \cite{34,35}. This method advantages a better control of the
cluster size but,
it must be pointed out that the obtaining of the negative metallic
ions is, generally, more difficult to do, this process requiring appropriate
conditions related to the external factors as temperature, electric field as
well as the filling factors; when large impurity concentration ($\approx
10^{18}$) is used then insignificant amount of the negative metallic 
centres are obtained.
In the
present paper, we show the experimental data for metallic clusters obtained
by thermal annealing of the $KCl:Ag^{-}$ samples. 

$KCl$ single crystals containing $Ag^{+}$ ions in a concentration of $5\cdot
10^{17}$ $ions/cm^3$ have been grown by the Kyropoulos method in air. 
Under electrolytical colouring
performed by an usual device in air at $573K$ and at $8000$ $V/cm$ 
we obtain samples containing $Ag^-$ negative metallic centres (see the 
initial (1) peak at 290 nm in the Fig. 2a)
as well as few small silver clusters (see the intial (1) peak at 380 nm in 
Figs. 
2a and b). Thermal annealing at a given temperature of the $KCl:Ag^{-}$ 
samples
leads, progressively, both to the decrease of the $Ag^-$ amount and to
the obtaining of the other, more and more,
clusters (see the rise trend of the 
absorption curves). A possible scenario for the conversion of the $Ag^-$ 
ions 
towards $Ag^o$ centres and/or cluster states begins with
$Ag^- + kT \leftrightarrow F + Ag^{o}_{i}$
(the $F$ means $F$ centre and the $Ag^{o}_{i}$ means an interstitial
silver atom) \cite{34}. The following step after the above reaction should 
be 
the precipitation of the silver atoms. 

  Also, during the thermal annealing, the clusters
have an increase trend of their sizes.
Experimentally, this fact can be observed by a change of the optical spectra;
the absorption maximum shifts, progressively, towards high wavelengths. 
In Figs. 2a the evolution of the optical spectra for the sample is shown
with respect to the time of annealing at $800K$; the first absorption maxima
are due to $Ag^-$ centres (290 nm), the second absorption maxima are due to 
silver clusters and the third, very slight peaks are due to the $F$ centres 
(550 nm).
Another set of optical spectra, corresponding to a thermal annealing at 
$920 K$, is shown in Fig. 2b. In this figure we have eliminated the 
absorption maxima corresponding to the $Ag^-$ centres.

As we can observe, besides the formation of the silver clusters, the thermal 
treatment of the samples containing $Ag^-$ leads to the appearance of the $F$
centres (the second peak in the right side of the figures). The $F$ centre 
peak
($550 nm$) has, as it is well known and as one can observe in the Figs. 2a 
and b,
no shift during the thermal annealing. 
In contrast with the $F$ centre behaviour, the absorption
maximum of the silver clusters moves, as we can see, towards the high 
wavelengths in . Curiously enough,
despite the fact that both the samples arise from the same $KCl : Ag^-$ 
crystal,
the thermal annealing at $920 K$ provides a better production and 
conservation of the $F$ centres (see Fig. 2b). However, a long time 
annealing
will lead to destroy all the $F$ centres.

>From the optical spectra, the cluster radii are determined using
electrodynamic ({\it Mie}) theory \cite{37}. Application of this theory to
large metal clusters is successful and a review of the method as well as
more complementary features are given in the book of {\it Vollmer \/} and 
{\it Kreibig \/} \cite{37}. The extinction cross section is given by 
\begin{equation}
\sigma_{ext} = \frac{2 \pi}{k^2} \sum_{L=1}^{\infty} \left( 2L + 1 \right)
Re \left(a_L + b_L \right), 
\end{equation}
where $k$ is the wavevector and $a_L$ and $b_L$ are coefficients containing
Bessel and Hankel functions which depend on the complex index of refraction
of the particle, the real index of refraction of the surrounding medium and
the size parameter $x = k \cdot R$. The $R $ is the cluster radius. For
clusters larger than about $10 nm$ the size dependence of the optical
spectra is an {\it extrinsic cluster size effect \/} \cite{37} due to
electrodynamics of the excitation which is governed only by the dimension of
the particle with respect to the wavelength of the light.

In Figs. 3a and b it is shown the time of the thermal treatment and the
corresponding increase of the cluster radii for the thermal annealing both
at $800 K$ and at $920 K$. The shapes of the curves are identically which
means that no influence of the $F$ centres on the cluster size evolutions
exists. Consequently, in the following we will discuss only a set of
experimental data ($920 K$). \topmargin=-0.5in \baselineskip=24pt

\section{Results.}

\indent As it is well-known one can distinguish two regimes in the cluster
growth process. The initial stage of cluster increase, following 
the formation of nuclei, is, in our case, due to the conversion of the 
negative metallic centres (see the decrease trend of the $Ag^{-}$ peak 
(270 nm)). Thus, the main stage
of the phase separation proceeds as a uniform growth of a 
number of precipitate particles from the supersaturated matrix. In this way,
during the thermal treatment, the cluster radius increases initially due to
the addition of particles coming from the source of the solute ions. When
this concentration decreased, the increase of cluster radius is due, mainly,
to particle transport from small clusters to larger ones. This is the second
stage in the cluster growth process which is known as the {\it Ostwald
Ripening (OR) \/}. It must be pointed out that the delimiting of the
boundary between the two increasing regimes is difficult to do, but one
supposes that the $OR$ regime begins usually before the solute concentration
decreases considerably \cite{29}. In Fig. 4 we have shown the theoretical 
curve
(equation 15) derived within the {\it off-centre diffusion \/} approach, the
theoretical curve corresponding with the $LSW$ theory and the experimental
curve for the increase of the cluster radius in the $OR$ stage for a 
thermal
annealing at $920K$. We have approximated the start of the $OR$ process
around of the $R=16nm$. For the theoretical calculus (the {\it off-centre
diffusion \/} approach) we have used appropriate values for dissociation
energy as $E_D=0.7eV$ and for the threshold energy of the particles transfer
as $E_b-e\Phi =1eV$. One can say that there is an agreement between the
theoretical curve derived within the {\it off-centre diffusion \/} approach
and experimental curves. This agreement becomes much better for larger radii
($R>18nm$). In contrast to the former, the agreement between the $LSW$
theory and the experiment is very good for $17nm<R<19nm$ and more poor in
the rest.

\topmargin=-0.5in \baselineskip=24pt

\section{Conclusion.}

\indent In summary, by the present paper, we have analyzed how the
correlation effects \cite{23,24} can be taken into account within an {\it 
off-centre diffusion approach \/} of the $OR$ process. The time dependence
of the cluster growth derived by this theoretical approach, under the
assumptions established in the introductory part, differs from that
predicted by the {\it LSW theory \/} but agrees with the most simulations
and experiments in the sense that $\alpha <\frac {1}{3}$ \cite
{9,10,11,12,13,14,15,16,17,18,19,20,21,22}. It must be pointed out that
though the dynamics is different from that predicted by $LSW$ theory and,
one seems that there is a better agreement of the theoretical approach
presented here with the experimental data for larger clusters ($R>18nm$),
this fact do not invalidate the known theoretical results \cite
{1,2,3,4,5,6,7,8}. A source of this difference could consist within the
stress effect in the host matrix and/or the correlation effects do not exist
in three-dimensional case. Consequently, the experimental data going over 
$R=20nm$ and a careful study of the stress effect should be helpful.

Beforehand to check again the agreement between the theory and the
experiment we can see that this approach is distinguished from the others in
this branch by the fact that the clusters act as individual entities and,
hereby, allowing of the introduction of the cluster properties according
with the recent discoveries \cite{31}: the cohesive energy, the dissociation
rate for the shrinking cluster, the mobility of the particles inside the
cluster which promotes the quasi-sphericity of the cluster shape, the
cluster kinks as the sites from where the particles leave the surface of the
shrinking cluster or the sites where these particles condense on the surface
of the growing cluster. However, despite of a relative agreement only
between theory and experiment below $R = 18 nm$ we may say that the results
are encouraging in further pursuing of this {\it off-centre diffusion
approach \/}. Also, we may say that a careful investigation on the transfer
frequency $p$ (within which the explicit form of the electrostatic potential
due to medium polarization should take into account) can rise the agreement
between the experimental and theoretical data.\\ {\it Acknowledgements. \/}
F. Despa is grateful to Professor M. Apostol for many useful discussions
regarding the theoretical results.

\newpage
\medskip
\centerline{\it Fig. 1 \/} \medskip
\noindent The way of the particles transport between the shrinking cluster
and the growing one. The particles leave the surface of the shrinking
cluster from the "kink positions" (the $N_o$ {\it off - centre sites \/})
and "condense" in the "kink positions" (the $N$ {\it off - centre sites \/})
of the growing cluster.

\medskip
\centerline{\it Fig. 2 a and b\/} \medskip
\noindent The changes of the absorption spectra during thermal treatment at $
800 K $ and at $920 K$. The second peak in (b) is due to the $F$ centres.

\medskip
\centerline{\it Fig. 3 a and b \/} \medskip
\noindent The increase of the cluster radius $R$ (\AA) versus the time $t$
(min.) of the thermal treatment at $800 K$ and at $920 K$.

\medskip
\centerline{\it Fig. 4 \/} \medskip
\noindent The time $t$ (min.) of thermal treatment at $920 K$ for the
increase of the $Ag$ cluster radii $R$ (\AA) at different sizes in the $OR$
stage.\\ $\Box $ - experimental,\\ -- - theoretical ({\it off centre
diffusion \/} approach),\\ $\dagger $ - theoretical ( $LSW$ theory).\\ The $
Ag$ clusters are embedded in the $KCl$ matrix.


\newpage

\begin{thebibliography}{99}
\bibitem{1}  Lifshitz M. and V. V. Slyozov: J. Phys. Chem. Solids {\bf 19},
35 (1961); Wagner C.: Z. Electrochem. {\bf 65}, 581 (1961).

\bibitem{2}  Rao M., M. H. Kalos, J. L. Lebowitz, and J. Marro: Phys. Rev. 
{\bf B13}, 4328 (1976).

\bibitem{3}  Lebowitz J.L., J. Marro, and M. H. Kalos: Acta Metall. {\bf 30}
, 297 (1982).

\bibitem{4}  Voorhees P.W. and M. E. Glicksman: Acta Metall. {\bf 32}, 2001
(1984).

\bibitem{5}  Grest G.S. and D. J. Srolowitz: Phys. Rev. {\bf B30}, 5150
(1984).

\bibitem{6}  Mazenko G.F., O. T. Valls, and F. C. Zhang: Phys. Rev. {\bf B31}
, 4453 (1985); {\bf 32}, 5507 (1985).

\bibitem{7}  Marder M.: Phys. Rev. Lett. {\bf 55}, 2953 (1985); {\it Phys.
Rev.} {\bf A36}, 858 (1987).

\bibitem{8}  Marqusee J.A. and J. Ross: J. Chem. Phys. {\bf 80}, 536 (1984).

\bibitem{9}  Langer J.S., M. Baron and H.D. Miller: Phys. Rev. {\bf A11},
1417, (1975).

\bibitem{10}  Binder K. and D. Stauffer: Phys. Rev. Lett. {\bf 33}, 1006,
(1974); K. Binder, D. Stauffer and H. Muller-Krumbhaar: Phys. Rev. {\bf B12}
, 5261, (1975).

\bibitem{11}  Mazenko G.F., O.T. Valls and F.C. Zhang: Phys. Rev. {\bf B31},
4453, (1985); {\bf 32}, 5507, (1985); G.F. Mazenko and O.T. Valls: Phys.
Rev. Lett. {\bf 59}, 680, (1987).

\bibitem{12}  Huse D.A.: Phys. Rev. {\bf B34}, 7845, (1986).

\bibitem{13}  Amar J.G., F.E. Sullivan and R.D. Mountain: Phys. Rev. {\bf B37},
196, (1988).

\bibitem{14}  Roland C. and M. Grant: Phys. Rev. Lett. {\bf 60}, 2657,
(1988).

\bibitem{15}  Gaulin B.D., S. Spooner and Y. Morii: Phys. Rev. Lett. {\bf 59},
668, (1987).

\bibitem{16}  Wiltzius P., F.S. Bates and W.R. Heffner: Phys. Rev. Lett. 
{\bf 60}, 3798, (1988).

\bibitem{17}  Raux D.: J. Phys. (Paris) {\bf 47}, 733, (1986).

\bibitem{18}  Binder K.: Phys. Rev. {\bf B15}, 4425, (1977).

\bibitem{19}  Schrobinger M., S.W. Koch and F. Abraham: J. Stat. Phys. {\bf 
42}, 1071, (1986).

\bibitem{20}  Kolb M., T. Groban, J.-F. Gouyet, and B. Sapoval: Europhys.
Lett. {\bf 11}, 601, (1990).

\bibitem{21}  Rogers T.M. and R. Desai: Phys. Rev. {\bf B39}, 11956, (1989).

\bibitem{22}  Bassereau P., D. Brodbreck, T.P. Russell, H.R. Brown and K.R.
Shull: Phys. Rev. Lett. {\bf 71}, 1716, (1993).

\bibitem{23}  Tokuyama M. and K. Kawasaki: Physica (Amsterdam) {\bf 123A},
386, (1984).

\bibitem{24}  Krichevsky O. and J. Stavans: Phys. Rev. Lett. {\bf 70}, 1473
(1993).

\bibitem{25}  Despa F.: Phys. Rev {\bf E} - to appear.

\bibitem{26}  Despa F. and M. Apostol: Solid State Commun. {\bf 94}, 153
(1995); for instance see also Despa F.: Phys. Stat. Sol.(b), {\bf 191}, 31
(1995).

\bibitem{27}  Despa F. and M. Apostol: J. of Phys. and Chem. Sol. - in print;
see also ICTP - Trieste preprint IC / 95 / 161.

\bibitem{28}  H$\ddot a$nggi P., P. Talkner and M. Borkovec: Rev. of Modern
Phys., {\bf 62}, 251 (1990).

\bibitem{29}  Hughes A. E. and S. C. Jain: Advances in Physics {\bf 28},
717, (1979); see also the references therein.

\bibitem{30}  Martin T.P., T. Bergmann, H. Gohlich and T. Lange: Z. Phys. 
{\bf D19}, 25 (1991).

\bibitem{31}  De Heer W.A.: Rev. of Modern Phys. {\bf 65}, 611, (1993).

\bibitem{32}  Jarrold M.F., U. Ray, J.E. Bower and K.M. Creegan: J. Chem.
Soc. Faraday Trans. {\bf 86 }, 2537, (1990).

\bibitem{33}  M. E. Davis and J. A. Mc Common, {\it Chem. Rev.} {\bf 90},
509 (1990).

\bibitem{34}  see for example the old papers: Kleeman W.: Z. Physik. {\bf 214},
 285 (1968); {\bf 215}, 113 (1968).

\bibitem{35} Topa V., A. Ioan, B. Iliescu and G. Mitroaica: Rev. Roum. Phys.
{\bf 18}, 571 (1973).

\bibitem{36}  Nistor L.C., V. Teodorescu, V. Topa, D. Topa and S.V. Nistor:
Cryst. Latt. and Amorph. Mat. {\bf 16}, 63 (1987).


\bibitem{37}  Vollmer M. and Kreibig U., {\it Optical Properties of Metal
Clusters\/}, Springer Ser. Mat. Sci., (1995); see also the references
therein.

\end{thebibliography}
\end{document}